\documentclass[10pt,twocolumn,aps,prb,amsmath,amssymb,a4paper,superscriptaddress]{revtex4-1}

\usepackage{graphicx, color,xkeyval,xparse,xstring,wasysym}
\usepackage[colorlinks=true,citecolor=black,linkcolor=black,urlcolor=blue,anchorcolor=black]{hyperref}
\usepackage[capitalise]{cleveref}  % must follow after pgfplots, pgfplotstable, or pdf
\usepackage[normalem]{ulem}

\definecolor{colspinup}{RGB}{215,48,39}
\definecolor{colspindown}{RGB}{69,117,180}

\DeclareDocumentCommand\opr{m}{\ensuremath{\mathbf{#1}}} % operator syntax
\DeclareDocumentCommand\Tr{r[]}{\ensuremath{\mathrm{Tr}\left[ {#1}\right]}} % trace syntax
\DeclareDocumentCommand\Imag{r[]}{\ensuremath{\mbox{Im}\left[ {#1}\right]}} % imaginary syntax
\DeclareDocumentCommand\Real{r[]}{\ensuremath{\mbox{Re}\left[ {#1}\right]}} % imaginary syntax
\DeclareDocumentCommand\log{o}{\ensuremath{\IfNoValueTF{#1}{\mbox{log}}{\mbox{log}\left({#1}\right)}}} % log fn syntax
\def\ct{\dagger}

\DeclareDocumentCommand\degrees{m}{${#1}^{\circ}$} % degrees syntax
\def\spup{\uparrow}
\def\spdn{\downarrow}

\def\unit#1{\,\textup{#1}}

\begin{document}
% ------------------------------------------------------------------------- %

\title{Charge and Spin Transport Anisotropy in Nanopatterned Graphene}

\author{S{\o}ren Schou Gregersen}
\email{sorgre@nanotech.dtu.dk}
\affiliation{Center for Nanostructured Graphene (CNG),
             DTU Nanotech, Department of Micro- and Nanotechnology,
             Technical University of Denmark, DK-2800 Kongens Lyngby, Denmark}
\affiliation{DTU Compute, Department of Applied Mathematics and Computer Science,
             Technical University of Denmark, DK-2800 Kongens Lyngby, Denmark}
\author{Jose H. Garcia}
\affiliation{Catalan Institute of Nanoscience and Nanotechnology (ICN2),
             CSIC and The Barcelona Institute of Science and Technology,
             Campus UAB, Bellaterra, 08193 Barcelona (Cerdanyola del Vall\`es), Spain}
\author{Antti-Pekka Jauho}
\affiliation{Center for Nanostructured Graphene (CNG),
             DTU Nanotech, Department of Micro- and Nanotechnology,
             Technical University of Denmark, DK-2800 Kongens Lyngby, Denmark}
\author{Stephan Roche}
\affiliation{Catalan Institute of Nanoscience and Nanotechnology (ICN2),
             CSIC and The Barcelona Institute of Science and Technology,
             Campus UAB, Bellaterra, 08193 Barcelona (Cerdanyola del Vall\`es), Spain}
\affiliation{ICREA, Instituci\'o Catalana de Recerca i Estudis Avan\c{c}ats,
             08070 Barcelona, Spain}
\author{Stephen R. Power}
\email{stephen.power@icn2.cat}
\affiliation{Catalan Institute of Nanoscience and Nanotechnology (ICN2),
             CSIC and The Barcelona Institute of Science and Technology,
             Campus UAB, Bellaterra, 08193 Barcelona (Cerdanyola del Vall\`es), Spain}
\affiliation{Universitat Aut\`onoma de Barcelona,
             08193 Bellaterra (Cerdanyola del Vall\`es), Spain}
\date{\today}

% ------------------------------------------------------------------------- %
\begin{abstract}
Anisotropic electronic transport is a possible route towards nanoscale circuitry design, particularly in two-dimensional materials. 
Proposals to introduce such a feature in patterned graphene have to date relied on large-scale structural inhomogeneities. 
Here we theoretically explore how a random, yet homogeneous, distribution of zigzag-edged triangular perforations can generate spatial anisotropies in both charge and spin transport.
Anisotropic electronic transport is found to persist under considerable disordering of the perforation edges, suggesting its viability under realistic experimental conditions.
Furthermore, controlling the relative orientation of perforations enables spin filtering of the transmitted electrons, resulting in a half-metallic anisotropic transport regime. 
Our findings point towards a co-integration of charge and spin control in a two-dimensional platform of relevance for nanocircuit design. 
We further highlight how geometrical effects allow finite samples to display finite transverse resistances, reminiscent of Spin Hall effects, in the absence of any bulk fingerprints of such mechanisms, and explore the underlying symmetries behind this behaviour.

\end{abstract}
% ------------------------------------------------------------------------- %

\maketitle

\section{Introduction}

An anisotropic electronic transport response in a system, where the ease with which electrons flow depends on the measurement direction, is an important and challenging concept with a number of key potential applications.
Such a behavior can be used to engineer electronic circuits and waveguides~\cite{Areshkin2007}, optical circuits~\cite{Kim2011}, and communications devices~\cite{Yan2012}.
It is particularly promising in two-dimensional materials, where the reduced dimensionality allows significant tuning of their electronic properties with only minor structural  modifications \cite{Pedersen2008,Yuan2013a,Pedersen2014, Power2014, Gregersen2016,Gregersen2017, son2017study}.
Indeed, much of the recent excitement surrounding monolayer black phosphorus, or phosphorene, relates to its intrinsic electronic anisotropy arising from a buckled geometry~\cite{Qiao2014,Liu2014}.
However, the high chemical reactivity of phosphorene presents a significant obstacle to its incorporation in devices\cite{PhysRevLett.114.046801, koenig2014electric}, and another strategy for implementing two-dimensional electronic anisotropy needs to be envisioned.
One possibility is to induce anisotropic behavior in an otherwise isotropic material.
Graphene is the most natural candidate for such an approach due to its exceptional electronic quality, ease of fabrication and electronic measurement, and a range of constantly improving patterning and etching techniques~\cite{CastroNeto2009a,Novoselov2012,Ferrari2015}.

Anisotropic transport has been experimentally demonstrated in graphene under strain~\cite{Pereira2009a,Kim2009}, in graphene sheets with periodic nanofacets~\cite{Odaka2010}, and has been suggested in anisotropically arranged graphene antidot lattices~\cite{Pedersen2014}. 
The latter case consists of an array of perforations, or \emph{antidots}, whose spacing is different along the $x$ and $y$ directions. 
All these methods introduce a system-wide physical anisotropy into the graphene sheet to break the equivalence of transport in the $x$ and $y$ directions.
In this work we consider an alternative patterning approach based on uniformly distributed perforations, where the anisotropic behavior is dictated by the atomic-level properties of the perforation edges.

Extended edges with the zigzag (zz) geometry locally break sublattice symmetry, leading to the formation of localized states~\cite{Nakada1996,Wang2016}, and to the formation of local magnetic moments when electron-electron interactions are considered~\cite{Son2006,Yu2008,Yazyev2010,Han2014,roche2015graphene}.
Local moments of opposite sign occur at the edges associated with different sublattices, so that global ferromagnetism can only occur when the overall sublattice symmetry is broken~\cite{Son2006,Wimmer2008,Saffarzadeh2011}, in accordance with Lieb's theorem~\cite{Lieb1989}.
This does not occur for zz-edged nanoribbons~\cite{Pisani2007} or perforations containing edges equally divided between the sublattices~\cite{Trolle2013}.
However, the three edges of a zz-edged triangular graphene antidot (zz-TGA) are all associated with a single sublattice, which dictates large ferromagnetic moments~\cite{Zheng2009b,Sheng2010,Potasz2012,Hong2016,Khan2016,Gregersen2016} (see Fig. \ref{fig:Geom}).

Recent works have demonstrated that lattices of approximately $1 \unit{nm}$ side length zz-TGAs provide an excellent platform for electronic and spintronic applications due to robust band gaps, half-metallicity, and spin-splitting properties\cite{Gregersen2016, Gregersen2017}.
In the case of spin-splitting, incoming currents can be directed into output leads according to their spin orientation.
Such behavior is analogous to the spin Hall effect\cite{Cresti2016}, but without relying on spin-orbit coupling (SOC) effects. 
These features were shown to be robust against disorder, unlike those in other antidot geometries, due to their dependence on local symmetry breaking effects and cumulative scattering from multiple antidots, and not on the exact separation and size of perforations. 
With state-of-the-art lithographic methods, triangular holes in graphene~\cite{Auton2016}, as well as zz-etched nanostructures~\cite{Shi2011,Oberhuber2013,Stehle2017}, can be realised.
Alternatively, a patterned layer of hexagonal boron nitride, into which triangular holes can be naturally etched with a large degree of control~\cite{Stehle2017}, could be employed as a lithographic mask. 
Many fabrication methods which give precise edge control require seeding, so that the exact distribution of features can be difficult to control.
Accounting for a random positioning of triangular defects is therefore crucial for realistic predictions in such systems. 
Recently, bottom-up techniques have also been developed which can independently give rise to perfect zigzag edges\cite{ruffieux2016surface} or perforations\cite{Moreno2018}, suggesting that further development in this field may allow a combination of these features in a given sample.
%We address this issue here by studying systems which no longer resemble regular superlattices.

Here we present simulations which demonstrate how the scattering properties of zz-TGAs also give rise to a significant transport anisotropy%
---namely a higher conductivity is observed for the armchair (ac) than the zigzag direction. 
Furthermore, inspired by the spin Hall-like transport behavior in finite TGA superlattice devices~\cite{Gregersen2017}, we also examine their bulk analog by calculating off-diagonal Hall conductivities.
Negligible values are obtained for both the charge and spin conductivities, contrasting with non-zero Hall-like transport in certain device geometries.
Using time-reversal symmetry (TRS) arguments, we reconcile these seemingly contradictory results and highlight the role they may play in experiment.
We consider zz-TGAs with side lengths $\sim \!5 \unit{nm}$, which are distributed randomly throughout the sample.
Two small sections of our simulated samples are depicted in \cref{fig:Geom}(a) and (b), with red and blue triangles denoting individual zz-TGAs.
With a homogeneous distribution (with respect to the $x$ and $y$ directions), anisotropic features originate solely from the edge structure of the zz-TGAs.
The strong positional disorder and broken translational invariance in all the simulated samples closely resembles what could be produced, for example, using seeded growth and selective edging techniques.
%In this way the simulated samples should more closely resemble what could be produced, for example, using seeded growth and selective edging techniques.
%All the simulated systems have extreme positional disorder, so that the material lack translational invariance.
Furthermore, we separately consider more experimentally relevant conditions by including significant side length and edge disorder, and comparing these systems to those with individually precise zz-TGAs.

Disordered systems, containing millions of atoms and hundreds of structural perforations, are simulated using a combination of mean-field Hubbard approaches for individual zz-TGAs and large-scale quantum transport methods to study  composite systems. 
The electronic behavior is analysed using the Chebyshev expansion of the density of states (DOS)~\cite{Weisse2006,Roche1999}, while the longitudinal~\cite{Roche1997,Lherbier2008,ROCHE20121404,Torres2014} and transverse~\cite{Garcia2015,Garcia2016,Cresti2016} conductivities are computed using the Kubo-Greenwood formalism. 
The Kubo method gives direct access to both the diagonal conductivities $\sigma_{xx}$ and $\sigma_{yy}$, from which the charge transport anisotropy $\sigma_{yy}/\sigma_{xx}$ can be measured, and the off-diagonal conductivities $\sigma_\textup{xy}$, which can be connected to measurements of the spin Hall effect.
Further details of the geometry and simulation methodologies are presented in Section \ref{sec:Geo}, after which we consider the most general case of randomly oriented triangles in \cref{sec:randomTGAs}.  Here we find that a significant anisotropy arises in the electronic transport, with identical behavior for each spin channel.
Upon considering triangles with the same alignment in \cref{sec:alignedTGAs}, a marked increase in the strength of the anisotropy is obtained, with the emergence of a robust half-metallicity, hence yielding  a spin-selective anisotropic transport behavior. In \cref{sec:spinisotropy}, we discuss the off-diagonal conductivity and reconcile the zero signal observed here with the previously predicted spin-Hall type behavior for zz-TGAs in a finite device geometry.
We conclude by summarising and discussing our findings in light of recent developments in the field, particularly those investigating spin transport mechanisms in graphene-based systems. 

\section{Geometry and model}
\label{sec:Geo}

\begin{figure}
\centering
\includegraphics{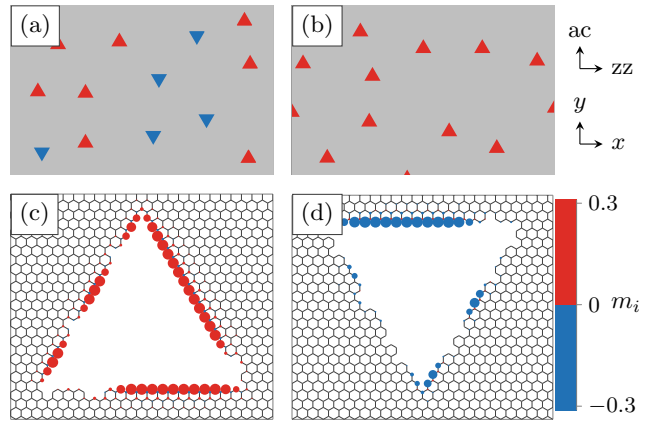}
\caption{
	Schematic illustrations of the distributions of a) randomly oriented and b) aligned zz-TGAs, together with the atomic structures and magnetic moment profiles of two edge-disordered TGAs with opposite alignment (c,d).
    In all panels, blue (red) colouring indicates spin-up (spin-down) polarization.
}
\label{fig:Geom}
\end{figure}

We consider samples of approximately $360 \unit{nm} \times 360 \unit{nm}$, which are periodic in both dimensions, and contain 400 randomly embedded zz-TGAs and almost 5 million atoms. 
Small sections from two of our samples are illustrated schematically in \cref{fig:Geom}(a) and (b).
The sample in \cref{fig:Geom}(a) illustrates the most general distribution, where not only the position, but also the orientation, of each triangle is randomised.
The sixfold symmetry of the graphene lattice allows for only two possible orientations of zz-TGAs.
Each orientation exposes zz-edges from different sublattices and, in turn, the two orientations exhibit magnetic moments of opposite sign.
We use red and blue coloring throughout this work to represent the spin up and down orientation, respectively.
The sample in \cref{fig:Geom}(b) shows the case where all the triangles are aligned and have the same magnetic orientation.
Note that the $x$ ($y$) axis in this work is aligned with a high symmetry zz (ac) direction of the underlying graphene lattice.

All samples discussed in this work contain randomly positioned zz-TGAs.
However, we also compare the cases when \emph{individual} perforations are either pristine or disordered by including side length variations and edge roughness.
In the pristine case, the zz-TGAs have side length $L=20 a$ ($\sim 5 \unit{nm}$), where the graphene lattice constant is $a = 2.46 \unit{Å}$, and in the disordered case we consider side lengths in the range $L = 22  \pm 8 a$.
This size allows us to determine if properties predicted in small systems persist to larger scales, whilst also allowing a large number of triangles to be included in the system.
Edge-disordered TGAs undergo a further simulated etching, where edge atoms are removed with probability $p=0.05$.
This etch is performed $4$ times.
Previous studies~\cite{Gregersen2017} indicate that side-length variation has only a minor impact on the qualitative transport-scattering mechanisms of small-scale zz-TGAs.
On the other hand, edge disorder has a more dramatic effect, reducing the corresponding magnetic moment distributions and affecting the spin-polarization properties.

The calculations are performed using a nearest-neighbor tight-binding model Hamiltonian
\begin{equation}
	\mathcal{H}_{\sigma} = \sum_{i} \epsilon_{i\sigma} {\opr c}_{i\sigma}^{\ct} {\opr c}_{i\sigma} + \sum_{ij} t_{ij} {\opr c}_{i\sigma}^{\ct} {\opr c}_{j\sigma} \,,
\end{equation}
where ${\opr c}_{i\sigma}^{\ct}$ (${\opr c}_{i\sigma}$) is the creation (annihilation) operator for spin $\sigma$ on site $i$.
The hopping parameter $t_{ij}$ is $ t = -2.7 \unit{eV}$ for neighbors $i$ and $j$, and zero otherwise.
No spin-orbit terms are included in the calculation.
Local magnetic moments are included via spin-dependent on-site energies $\epsilon_{i\sigma} = \pm \frac{U}{2} m_i$, with $-$ for $\sigma=\spup$ and $+$ for $\sigma=\spdn$.
The on-site magnetic moments $m_i$ are calculated from a self-consistent solution of the Hubbard model within the mean field approximation, $m_i=\left<{\opr n}_{i\spup}\right> - \left<{\opr n}_{i\spdn}\right>$, where ${\opr n}_{i\sigma}$ is the number operator.
The magnetic moment distributions for individual triangles are calculated in periodic systems using $50 a \times 30 \sqrt{3} a$ unit cells (approximately $12 \unit{nm} \times 12 \unit{nm}$), which are then embedded into the larger samples.
A minimum separation of $X_\text{min} \approx 12 \unit{nm}$ is imposed between neighboring triangles, which prevents the unit cells used in the mean-field parameterization of individual perforations from overlapping.
The on-site Hubbard parameter is set as $U= 1.33 |t|$, which gives good agreement with \emph{ab initio} calculations in the case of graphene nanoribbons~\cite{Yazyev2010}.

The atomic structure of two such edge-disordered zz-TGAs and their associated magnetic moments are displayed in \cref{fig:Geom}(c) and (d).
Note that the orientations of the TGAs in \cref{fig:Geom}(c) and (d) are rotated \degrees{60} with respect to one another, which in turns yields oppositely spin-polarized edges.
While in one case, \cref{fig:Geom}(c), all three edges still display significant magnetic moments, in the other case, \cref{fig:Geom}(d), significantly lowered magnetic moments are seen.
A small number of TGAs in our samples will have a considerable quenching of the magnetic moments along one or more of their edges.

The DOS is determined using the efficient linear-scaling kernel polynomial method~\cite{Weisse2006} with the Jackson Kernel, and a spectral resolution of approximately $4 \unit{meV}$.
The conductivity tensor $\sigma_{\alpha,\beta}$ is determined within the linear response regime, using the Kubo-Bastin formula\cite{Bastin1971}:
\begin{equation}
\sigma_{\alpha\beta}^\eta = \frac{ie^2}{\Omega}\hbar\int_{-\infty}^E \text{Tr}\left[v_\alpha\, \frac{dG^+(H,\varepsilon)}{d\varepsilon}\, v_\beta\,\delta(H-\varepsilon) - h.c \right] d\varepsilon,  \label{Kubo-Bastin}
\end{equation}
where E the Fermi energy, $G^+(H,\varepsilon)$ the advanced Green's function, $\delta(x)$ the Dirac's delta function and $v_\alpha\equiv [H,X_\alpha]/i\hbar$ the $\alpha$-component of the velocity operator.
The off-diagonal elements are computed numerically by using an efficient linear-scaling algorithm based on the kernel polynomial method~\cite{Garcia2015, Cresti2016}. 
For the diagonal elements, Eq. (\ref{Kubo-Bastin}) simplifies to the Kubo-Greenwood formula, where a more efficient real-space time-dependent wavepacket propagation method~\cite{Roche1997,Lherbier2008,ROCHE20121404,Torres2014} is applicable.
In this approach, a random wavepacket is injected into the system.
In the presence of disorder, the propagation of the packet quickly becomes diffusive.
In such situations, one can relate the time evolution of the mean square displacement operator $\Delta X(E,t)$\cite{Torres2014} to the diffusive coefficient 
\begin{equation}
D(E, t) = \frac{\partial}{\partial t }\Delta X(E,t)
\end{equation}
which can finally be used to compute the conductivity by the means of Einstein's relation
\begin{equation}
\sigma_{\alpha\alpha}(E)=\lim_{t\rightarrow \infty}\sigma_{\alpha\alpha} (E, t)=\lim_{t\rightarrow \infty} \frac{1}{2}e^2\rho(E)D(E,t).
\end{equation}
where $\rho(E)$ the density of states, and $\sigma_{\alpha\alpha} (E, t)$ is a timescale-dependent conductivity.
The semiclassical conductivity $\sigma_{\text{sc}}$ is obtained when the diffusion coefficient reaches a saturation limit (maximum value), where $\sigma_{\text{sc}}\equiv \frac{1}{2} e^2 \rho(E) D_\text{max}(E,t)$, whereas at longer times localization effects start to dominate and suppress the conductivity, eventually leading to an insulating behavior.
This method also allow for exploring both limits by exploiting the relation between the effective system size $L$ and the mean square displacement $L\equiv\sqrt{\Delta X^2(E,t)}$, which permits us to select sizes that are much larger than the mean free path but shorter than the localization length\cite{Torres2014}.

\section{Randomly oriented triangles}
\label{sec:randomTGAs}

\begin{figure}
\centering
\includegraphics{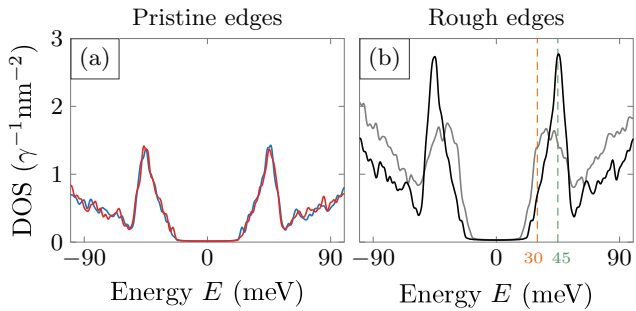}
\caption{
    Density of states (DOS) for systems with randomly oriented triangles, as illustrated in \cref{fig:Geom}(a).
    Panel (a) shows the spin-dependent DOS for each spin orientation (blue and red curves) for the case with pristine edges and panel (b) compares the total DOS for the pristine (black) and disordered (gray) edges.
    A prominent band gap is seen in all cases.
    The energies $30, 45 \unit{meV}$ are marked for later discussion.
}
\label{fig:flip.DOS}
\end{figure}

We first consider the case illustrated in \cref{fig:Geom}(a), where both the position and orientation of the triangles are randomized.
Each (spin) orientation occurs with the same probability and, as expected, we find no significant spin-dependence in either the DOS or transport results.
For the case without edge disorder, the spin-up and spin-down DOS (red and blue curves, respectively) in \cref{fig:flip.DOS}(a) are almost exactly superimposed, and they have half the value of the total DOS (black curve) displayed in \cref{fig:flip.DOS}(b).
Minor differences between the spin channels become negligible as the system size is further increased.
The system displays semi-conducting behavior, with a gap of approximately $40 \unit{meV}$.
Similar band gap and spin-unpolarized behavior has been demonstrated for periodic superlattices~\cite{Gregersen2016}, but emerges here for a completely random distribution of antidots.
Including edge-disorder, as in the gray curve of \cref{fig:flip.DOS}(b), results in a qualitatively similar total DOS (dashed curve) with only a slight reduction of the semiconducting gap.
This extraordinary robustness against disorder is in stark contrast to, for example, the band-gap formation in lattices of circular or hexagonal antidots, which is very sensitive to fluctuations in the lattice periodicity or antidot shape~\cite{Yuan2013a,Power2014}.
This robust band gap was previously studied for small-sized TGA superlattices~\cite{Gregersen2016}, but here it survives even without the constraint of a superlattice.
In \cref{fig:flip.DOS}(b) we highlight the energies $E= 30 \unit{meV}$ (orange) and $45 \unit{meV}$ (green), near the band-edge and peak of the non-zero DOS region respectively, which will be considered in more detail below. 

\begin{figure}
    \centering
\includegraphics{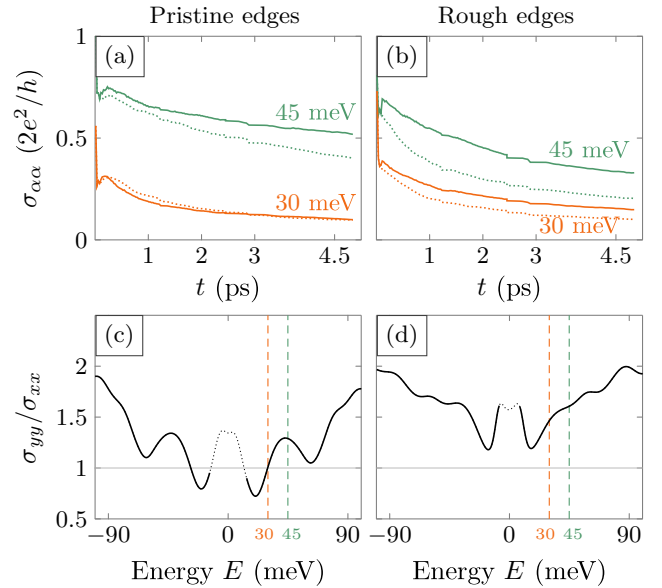}
    \caption{
        Conductivities $\sigma_{\alpha\alpha}$ (a,b) and anisotropies $\sigma_{yy} / \sigma_{xx}$ at the maximum simulation time $t_\textup{max} \approx 4.5 \unit{ps}$ (c,d) for a sample with randomly aligned triangles [see \cref{fig:Geom}(a)].
        The sample in (a) and (c) has pristine-edged triangles, and that in (b) and (d) includes edge disorder.
        The solid and dotted plots in (a,b) correspond to ac (y) and zz (x) directions, respectively.
        The horizontal gray line notes the isotropic case $\sigma_{yy} / \sigma_{xx} = 1$.
        In panels (c,d), the shaded region denotes energy ranges where the anisotropy is ill-defined.
        The energies $30,45 \unit{meV}$ are marked as in \cref{fig:flip.DOS}(b).
    }
    \label{fig:flip.sigma}
\end{figure}

The electronic transport for randomly-oriented zz-TGA systems is examined for both pristine and rough-edged triangles.
The time-dependent conductivity $\sigma_{\alpha\alpha}(E,t)$ gives access to the different transport regimes emerging during the time evolution.
Since both spins contribute equally, we consider only the total (charge) conductivity $\sigma = \sigma^{\spup} + \sigma^{\spdn}$.
The $x$- (dotted) and $y$- (solid) direction longitudinal conductivities for are given in the top panels of \cref{fig:flip.sigma} for $E= 30 \unit{meV}$ and $45 \unit{meV}$.
In all cases the conductivities reach a maximum value at short times $t < 0.5 \unit{ps}$, implying the onset of a diffusive regime, whereas localization effects enter into play at longer times.~\cite{Lherbier2008,Ortmann2011} 
We note that the graphene sections between perforations in our systems are pristine, so that at extremely short time scales the wave packet can explore this region with an extremely high conductivity, seen as sharp peaks near $t=0$ in our simulations. 
We neglect this regime throughout our discussion, which focuses on the effects introduced by the zz-TGAs.
The emergence of a transport anisotropy is clear by comparing the solid and dashed lines for the same energy, and it is particularly significant for the higher energy (green) case in the presence of disordered edges (\cref{fig:flip.sigma}(b)).
The anisotropies $\sigma_{yy}/\sigma_{xx}$ at the maximum simulation time $t_\textup{max} \approx 4.5 \unit{ps}$ are displayed in \cref{fig:flip.sigma}(c) and \cref{fig:flip.sigma}(d), for the pristine and disordered cases respectively, as a function of energy.
The faded sections of the curves correspond to the gapped regions of the DOS [see \cref{fig:flip.DOS}(b)], where the individual conductivities $\sigma_{yy}$ and $\sigma_{xx}$ are nearly zero, and the anisotropy is not well-defined.

In both \cref{fig:flip.sigma}(c) and (d) the transport anisotropy tends towards $\sigma_{yy}/\sigma_{xx} ~\sim 2$ at energies far from the gap, and decreases towards unity at the band edges near the gap.
Generally, the anisotropy lies above unity ($\sigma_{yy}/\sigma_{xx} > 1$), signifying a preferred ac ($y$) transport direction.
Electronic states near the band edges in zz-TGA systems are primarily localized near the edges of the TGAs, whereas states further inside the continuum tend to be more homogeneously dispersed~\cite{Gregersen2016}.
Away from the gap, the anisotropy can be explained by a scattering of these more dispersive states by the TGAs which is stronger for 
currents along the zz-direction than those in the ac-direction.
Near the band edges, though, the electronic states are less dispersive and display a lower conductivity.
Furthermore, the current in this case is carried by highly localized states near the triangular edges, so that pristine-edged triangles present very little scattering and result in a nearly isotropic transport regime.
It is interesting to note that the anisotropy is, in fact, larger for the sample which \emph{includes} edge disorder.
Disordered edges degrade transport more along the zz ($x$) directions, because this direction aligns with long nanoribbon-like edges, whose transmission channels are reduced significantly by disorder.
Edge-disordered samples are inevitable in realistic systems, where position, orientation, and edge roughness will be difficult to control.
It is reassuring that a homogeneous distribution of zz-TGAs, with random orientations and edge disorder, displays such a strong transport anisotropy.

\section{Half-metallic aligned samples}
\label{sec:alignedTGAs}

\begin{figure}
\centering
\includegraphics{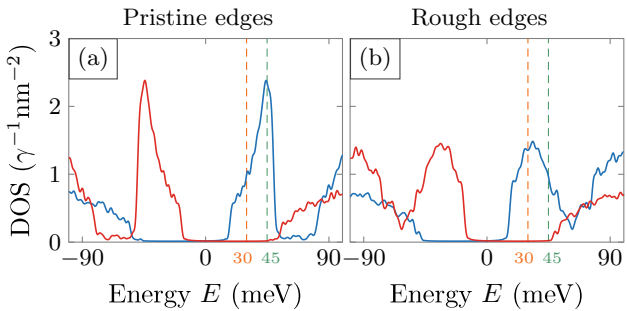}
\caption{
    Spin-dependent DOS (a) without edge disorder and (b) including edge disorder for the case of aligned TGAs as illustrated in \cref{fig:Geom}(b).
    The spin-up (down) polarization is shown by the red (blue) curve.
    The energies $30,45 \unit{meV}$ are marked as in \cref{fig:flip.DOS}(b).
}
\label{fig:algn.DOS}
\end{figure}

We now consider systems with all triangles aligned in the same direction, so only edges from one sublattice are exposed and all triangles have magnetic moments of the same orientation, as in \cref{fig:Geom}(b).
The spin-polarized DOS for pristine and disordered edges are shown in \cref{fig:algn.DOS}(a) and (b), respectively, where spin-up and spin-down DOS are displayed by separate red and blue curves.
An almost perfect spin polarized DOS is observed within $15 \unit{meV} \apprle |E| \apprle 50 \unit{meV}$, with opposite spin polarizations on either side of $E=0$.
These half-metallic regions are separated by a band gap of approximately $30 \unit{meV}$.
Just as in the case of randomly-oriented triangles, the band gap and spin-polarization trends are similar to previous results in superlattices.\cite{Gregersen2016}
The inclusion of edge defects does not affect the qualitative behavior, i.e. we still find two oppositely spin-polarized, half-metallic regions separated by a small band gap.
The principal effect of edge disorder is a slight smearing of the DOS features.
It is worth emphasizing that the near perfect half metallicity is exceedingly robust considering that it emerges from a random distribution of edge-disordered antidots.

\begin{figure}
\centering
\includegraphics{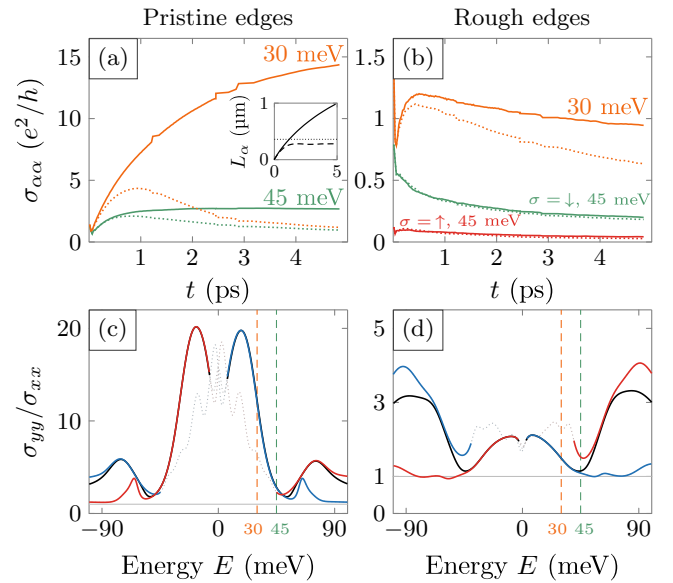}
\caption{
    The spin down conductivities $\sigma_{\alpha\alpha}$ (a,b) and anisotropies $\sigma_{yy} / \sigma_{xx}$ of both spin-types at the maximum simulation time $t_\textup{max} \approx 5 \unit{ps}$ (c,d) for the aligned triangles sample shown in \cref{fig:Geom}(b).
    The left (right) panels show cases with pristine (disordered) triangle edges. 
    Solid (dotted) curve shows the zz (ac) direction conductivity in panels (a,b).
    The horizontal gray line notes the isotropic case $\sigma_{yy} / \sigma_{xx} = 1$.
    Faded sections in panels (b,d) correspond to regions with ill-defined anisotropy. 
    The energies $30,45 \unit{meV}$ are marked as in \cref{fig:flip.DOS}(b).
    The inset in (a) displays the $E=30 \unit{meV}$ simulation lengths $L_\alpha$ as a function of time.
}
\label{fig:algn.sigma}
\end{figure}

Electronic transport simulations, without edge-disorder, reveal a large anisotropy, illustrated by the spin-down conductivities at the energies $E= 30 \unit{meV}$ and $45 \unit{meV}$ in \cref{fig:algn.sigma}(a).
The spin-up conductivity (not shown) is mostly zero at these energies due to half-metallicity.
The transport anisotropy is particularly large for the $E= 30 \unit{meV}$ case.
Here the $\sigma_{yy}$ (ac-direction, solid orange curve) shows quasi-ballistic behavior, i.e. a sub-linear increase, and, no sign of saturation within the accessible time range of 5 ps.
Meanwhile $\sigma_{yy}$ for the $E = 45 \unit{meV}$ case saturates to a diffusive, constant value.
For both energies $\sigma_{xx}$ localizes quickly at $t \sim 1 \unit{ps}$.
The simulations are stopped at 5 ps, at which time the simulation length $L_{y}$ for the $E=30\unit{meV}$ case (solid curve in the inset of \cref{fig:algn.sigma}(a)), exceeds the sample size $L_{y} > 360 \unit{nm}$.
When the simulation wave-packets become larger than the simulation samples, artifacts may develop originating solely from the periodic boundary conditions used.
The remaining conductivities in \cref{fig:algn.sigma}(a) have corresponding wavepacket propagation lengths below $360 \unit{nm}$ (the $x$-direction case at $E=30\unit{meV}$ is shown by the dashed line in the inset).

The conductivity anisotropies $\sigma_{yy}/\sigma_{xx}$ of the sample without edge-disorder, at the maximum simulation time $t_\textup{max} \approx 5 \unit{ps}$, are displayed in \cref{fig:algn.sigma}(c).
The spin-up and spin-down anisotropies are displayed by separate red and blue curves, with the total charge anisotropy shown in black.
The individual anisotropies are ill-defined in the dotted sections where conductivities $\sigma_{yy}$ and $\sigma_{xx}$ vanish.
The anti-symmetry in \cref{fig:algn.sigma}(c) between the two spin channels originates from the anti-symmetric DOS [see \cref{fig:algn.DOS}(a)].
\cref{fig:algn.sigma}(c) reveals a maximum anisotropy (for spin-down) near $E \sim \pm 20 \unit{meV}$, with $\sigma_{yy}/\sigma_{xx} \sim 20$.
This corresponds to the onset of states which are largely localized on the zz-edges, but have some penetration into the surrounding graphene~\cite{Gregersen2016}.
A comparison with the more isotropic transport found for random TGA orientations [see \cref{fig:flip.sigma}(c)] suggests that localised states at TGAs with the \emph{same} orientation couple quite strongly to form robust transport channels along the ac direction.
The anisotropy for spin-down decreases towards larger energies, but never reaches values below unity, and transport in the ac direction is always preferred.
At higher energies still, the onset of spin-up states (red curve, see also \cref{fig:algn.DOS}(a)) of a similar, but more dispersive, nature than their lower-energy spin-down counterparts gives a less massive, but still significant, anisotropy $\sigma_{yy}/\sigma_{xx} \sim 5$.
The total charge anisotropy (black curve) is lowest at the crossover between the spin-down and spin-up regimes.
Most importantly, since low-energy transport is driven by coupling between states localised near zz-TGAs, and not bulk-states which can be scattered by them, a quasi-ballistic regime emerges in which extremely large anisotropies are present.

The decrease in conductivity on including edge disorder is most evident for the $E = 30 \unit{meV}$ case (orange curves) in \cref{fig:algn.sigma}(b).
Now both $\sigma_{yy}$ (solid) and $\sigma_{xx}$ (dotted) saturate near $t \sim 0.5 \unit{ps}$ and afterwards begin to localize.
%Lower conductivities are found when edge disorder is included, as shown for $E = 30 \unit{meV}$, $45 \unit{meV}$, and $50 \unit{meV}$ in \cref{fig:algn.sigma}(c).
%In the $E= 30 \unit{meV}$ case, both $\sigma_{yy}$ (ac, solid) and $\sigma_{xx}$ (zz, dashed) saturate near $t \sim 0.5 \unit{ps}$ and afterwards begin to localize.
The $E=45 \unit{meV}$ conductivities saturate even earlier, and the maxima cannot be resolved.
All simulation lengths now remain within the sample size (not shown). 
Earlier onset of localized behavior was also observed in \cref{fig:flip.sigma}(b) for randomly-aligned triangles, and can be attributed to a degradation of transport mediated by channels near the triangle edges by scattering due to edge roughness.
The anisotropies $\sigma_{yy}/\sigma_{xx}$ of the edge-disordered sample are displayed in \cref{fig:algn.sigma}(d) at the maximum simulation time $t_\textup{max}$.
These show a continued preference for ac-direction transmission, as they again remain above unity.
However, the maximum is much smaller than the pristine case a reduction to $\sigma_{yy}/\sigma_{xx} \sim 2$ for the spin-down channel.
However, the higher-energy up-spin transport suffers a less dramatic reduction and takes values $\sigma_{yy}/\sigma_{xx} \sim 4$ at $E = \pm 90 \unit{meV}$.
The rough edges significantly quench electronic transport near the band edges, although not enough to completely remove the transport anisotropy.
However, the higher-energy states are more dispersed and thus less sensitive to edge roughness.
We note that controlling the orientation of a random distribution of zz-TGAs allows for behavior with huge potential for spintronic applications -- namely a device whose spin-polarization and degree of anisotropy is tunable by gate voltage. 
Furthermore, this behavior remains qualitatively similar under reasonable disorder of the triangle edges.
Considering only the total charge current (black curve in \cref{fig:algn.sigma}( d)), we note that aligned triangles give a similar anisotropy to the randomly aligned case, but with a somewhat greater magnitude.
This suggests that such setups may be interesting even for electronic devices which neglect the spin properties underpinning their anisotropic transport behavior.

\section{Off-diagonal conductivities} \label{sec:spinisotropy}

The anisotropic transport discussed so far emerges due to a geometric asymmetry---%
the $x$- and $y$-directions in the lattice have a different alignment relative to the zigzag edge segments.
A strong anisotropy then arises when transport is mediated by electronic states associated with these edges.
Edge magnetism can then further lead to a strong spin-dependent behavior.

\begin{figure*}
      \centering
\includegraphics{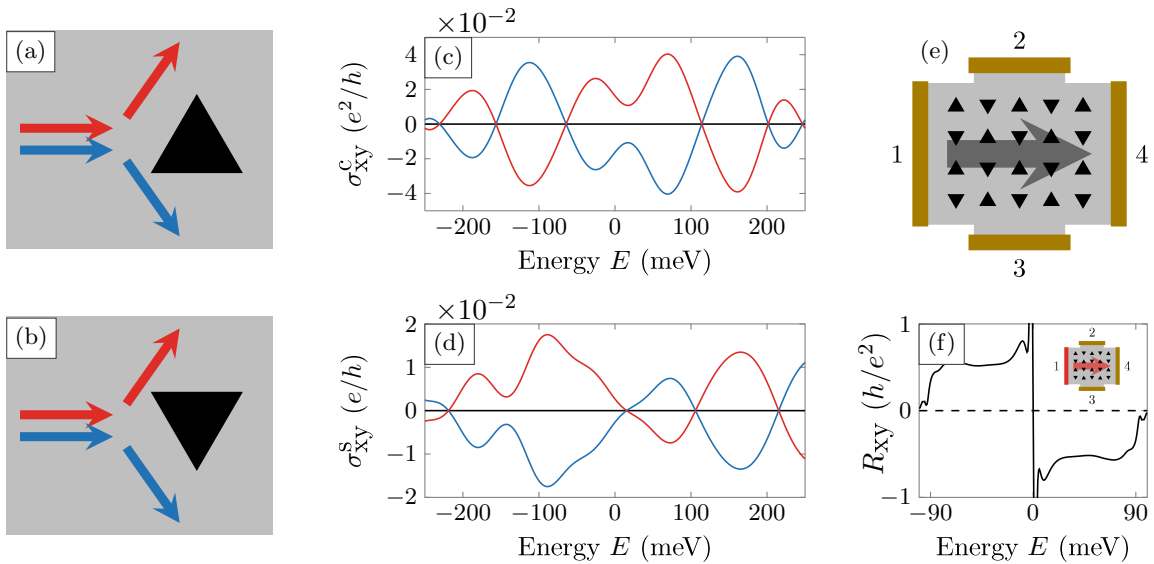}
      \caption{
            Spin scattering features (a,b), transverse conductivities in bulk (c,b), finite device schematic for separating spin types (e), and transverse spin-Hall-like resistances for said device (f).
            (a) and (b) The spin scattering around a zz-TGA, with the TGA in (a) rotated with respect to (b).
            Notice the spin-dependent scattering directions are \emph{the same} in (a) and (b), regardsless of TGA orientation.
            (c) and (d) The Hall and spin-Hall conductivities of the sample in \cref{fig:Geom}(a). The blue curves show $\sigma_{xy}(E)$ evaluated for a single random vector, and the red curves show $\sigma_{yx}(-E)$. These quantities should be equal by the Onsager relations, which indicate that the true Hall and spin-Hall conductivities are identically zero (black lines).            
            (e) Schematic illustration of device; a graphene cross bar with a $4 \times 5$ array of zz-TGAs of either orientation and correspondingly spin-polarized, originally proposed in previous work~\cite{Gregersen2017}.
            (f) Transverse (Landauer-B\"uttiker) resistances $R_\textup{xy}$ of the device illustrated in (e), with either spin unpolarized (dashed) or spin-up polarized (solid) left lead (as in inset).
      }
      \label{fig:spinsplitting}
\end{figure*}

A similar spin-dependent geometric asymmetry can be observed in the electronic scattering profile of zz-TGAs.
Electrons incident along the $x$-direction display radically different scattering probabilities in the $+y$ or $-y$ directions due to the broken reflection symmetry inherent in the triangular shape.
In a previous work~\cite{Gregersen2017}, we demonstrated how zz-TGA devices may be engineered to exploit this property to spatially split spins according to their orientation, as shown in \cref{fig:spinsplitting}(a) and (b).
In certain geometries, non-zero transverse resistances could be induced.
These behaviors suggest analogies with the Spin Hall and inverse Spin Hall effects (SHE/iSHE), which induce similar results when spin-orbit coupling (SOC) terms are included.

%The generation of transverse spin currents by SOC mechanisms has been investigated using both device (Landauer-Buttiker) and bulk (Kubo-Bastin) methodologies.
%It is worth 
%\begin{itemize}
%	\item exploring whether an extension of our bulk study to off-diagonal conductivities can further clarify the spin-current generation mechanisms at play in zz-TGA setups
%	\item discussing the different symmetry considerations that arise in zz-TGA and SOC systems, and
%	\item explicitly determining the requirements for a non-zero transverse charge or spin signal to be measured in a finite device geometry, and how these relate to bulk conductivity values.
%\end{itemize}

%\textcolor{red}{(What follows needs to be considerably condensed and/or some moved to appendix):}

These effects can be simulated in both finite device geometries, using Landauer-B\"uttiker methods, and in bulk systems, using the Kubo-Bastin approach.
Transverse resistances and spin currents, corresponding directly to experimentally-measurable quantities, emerge from multi-terminal simulations and are associated with the off-diagonal elements of the bulk conductivity tensor, $\sigma_{xy}$. 
This quantity captures the intrinsic response of the system, and unlike the Hall resistance $R_{xy}$, it is independent of the measurement geometry or device setup.
The off-diagonal matrix elements of the conductivity ($\sigma_{xy}$) and resistivity ($\rho_{xy}$) tensors are connected by 
\begin{equation}
	\rho_{xy} = \frac{\sigma_{xy}}{\sigma_{xx}\sigma_{yy} + \sigma_{xy}^2}\,
\end{equation}
from which we note that their zeros coincide in conducting systems. 
The Onsager reciprocity relation forces the off-diagonal elements of the conductivity tensor to be zero unless time-reversal symmetry (TRS) is broken.
However, even in cases when TRS is preserved, the off-diagonal tensor elements are not necessarily trivial once we consider individual spin channels.
Treating the spin channels as separate, non-interacting systems, SOC enters as an effective magnetic field which breaks TRS in each channel independently, giving finite $\sigma_{xy}^\uparrow$ and $\sigma_{xy}^\downarrow$.
However, the two spin channels form a time-reversal symmetric pair and contribute equally but oppositely to the off-diagonal conductivity, \emph{i.e.} $\sigma_{xy}^\uparrow = -\sigma_{xy}^\downarrow \ne 0 $.
The total, or \emph{charge}, off-diagonal conductivity $\sigma_{xy} = \sigma_{xy}^\uparrow + \sigma_{xy}^\downarrow$ is then zero, consistent with the conservation of TRS in the system as a whole.
However, the \emph{spin} Hall conductivity $\sigma_{xy}^{S} = \sigma_{xy}^\uparrow - \sigma_{xy}^\downarrow$ is finite.

%In the SOC case, the two spin channels form a time-reversal symmetric pair and contribute equally but oppositely to the off-diagonal conductivity, \emph{i.e.} $\sigma_{xy}^\uparrow = -\sigma_{xy}^\downarrow \ne 0 $ .
%The total, or \emph{charge}, off-diagonal conductivity $\sigma_{xy} = \sigma_{xy}^\uparrow + \sigma_{xy}^\downarrow$ is then zero, but the \emph{spin Hall} conductivity $\sigma_{xy}^{S} = \sigma_{xy}^\uparrow - \sigma_{xy}^\downarrow$ is finite.
%Treating the spin channels as separate non-interacting systems, SOC enters as an effective magnetic field which breaks TRS in each channel independently, giving finite $\sigma_{xy}^\uparrow$ and $\sigma_{xy}^\downarrow$, but conserves TRS for the complete system, so that $\sigma_{xy}$ = 0.

In the case of zz-TGA structures, the presence of magnetic moments breaks TRS.
However, the absence of spin-mixing terms allows the system to once more be considered as two separate, non-interacting spin channels.
Our model considers two parallel channels with different (i.e. spin-dependent) potential landscapes, and within each channel, TRS is effectively preserved (up to spin orientation) and thus the individual 
off-diagonal conductivities  $\sigma_{xy}^\uparrow$ and $\sigma_{xy}^\downarrow$ are identically zero.
The total off-diagonal conductivity $\sigma_{xy}$, being a sum of the two spin channel contributions, is also zero despite the broken time-reversal symmetry of the complete system.
A full Kubo-Bastin calculation confirms that both the Hall and Spin-Hall conductivities are zero, as shown by the solid black lines in \cref{fig:spinsplitting}(c) and (d), respectively. 
However, it is worth pointing out that the stochastic scheme used to compute both conductivities inherently breaks all the system's symmetries because it connects all its possible subspaces.
These symmetries need later to be restored, either by averaging over many random vectors, or by an explicit imposition of the symmetries through a set transformation which cancels additional random noise\cite{Marmolejo-Tejada2017}.
In \cref{fig:spinsplitting}(c,d), we show the result for an arbitrary random vector (blue curve) and note the non-zero, oscillatory behaviour.
A large set of random vectors is required for the averaging approach, and the result can display misleading behaviour en route to convergence.
Unlike the case of true spin or valley hall effects, we cannot take advantage of an overall TRS in the set transformation approach\cite{Marmolejo-Tejada2017}.
However, the spin-asymmetry around $E=0$ allows to write an Onsager relation\cite{Hertel2012} $\sigma_{xy}(E) = \sigma_{yx}(-E )$, and calculate $\sigma_{xy}(E)$ as an average of these two quantities evaluated for a small number of random vectors.
$\sigma_{yx}(-E )$ is plotted by a red curve in \cref{fig:spinsplitting}(c,d), and exactly cancels $\sigma_{xy}(E )$ for the same random vector, indicating that the charge \emph{and} spin Hall conductivities are identically zero.

In finite, multi-terminal devices, the Landauer formula
\begin{equation}
	I_{\alpha} = \frac{e^2}{h}  \sum_{\beta} \left( T_{\beta \alpha}  V_{\alpha} - T_{ \alpha \beta}  V_{\beta}\right) \,, 
	\label{eq:LB}
\end{equation}
can be used to calculate the currents ($I_{\alpha}$) and potentials ($V_{\alpha}$) at each lead from the transmissions ($T_{\beta \alpha}$) between leads.
To evaluate the transverse resistance in the four-terminal cross-bar shown in \cref{fig:spinsplitting}(e), we solve this system of equations under the boundary conditions $I_1 = -I_4 \equiv I$, $I_2=I_3=0$, $V_1\equiv V$, $V_4=0$, so that
\begin{equation}
	R_{xy} = \frac{V_{2} - V_3}{I} \equiv \frac{V_{23}}{I} \,.
\end{equation}
%and calculated from the potential difference induced between the top and bottom probes by a current driven from the left to right.
%From the Landauer formula
%\begin{align}
%	I_{\alpha} = \frac{e^2}{h}  \sum_{\beta} \left( T_{\beta \alpha}  V_{\alpha} - T_{ \alpha \beta}  V_{\beta}\right) \,, 
%	\label{eq:LB}
%\end{align}
%relating the currents in each probe ($I_{\alpha}$) to the transmissions $T_{\beta \alpha}$ between probes $\alpha$ and $\beta$, with an appropriate set of boundary conditions (current in = current out, probes 2 and 3 are floating voltage probes), we find in its most general form that
Restricting our analysis only to the conditions that give rise to a non-zero transverse resistance, we assume from here on that $R_{xy}$ follows $V_{23}$, and omit complicating prefactors and denominators.
In its most general form\cite{buttikersym}, 
\begin{equation}
\begin{aligned}
R_{xy} \sim  & \; (T_{13}T_{21} - T_{31}T_{12}) + (T_{31} +T_{21})(T_{23} - T_{32}) \\  + \, &  \; T_{21} T_{43} - T_{31} T_{42} \,,
\end{aligned}
\label{Rxy_full}
\end{equation}
and for independent spin channels, each transmission term can be written  $T_{\beta \alpha} = T_{\beta \alpha}^{\uparrow} + T_{\beta \alpha}^{\downarrow} $. 
Under TRS ($T_{\beta \alpha} =T_{\alpha \beta} $), Eq. \eqref{Rxy_full} simplifies to
\begin{equation}
	R_{xy} \sim T_{21} T_{43} - T_{31} T_{42} \,,
	\label{Rxy_simple}
\end{equation}
showing that the transverse resistance can be finite even when the associated resistivity is zero.
However, this discrepancy requires asymmetric couplings between sets of probes which we would generally expect to be symmetric (e.g. $T_{21}$ and $T_{24}$)  
%we require differences between some of the transmissions $T_{21}$, $T_{31}$, $T_{43}$ and $T_{42} $, each of which couple a current probe (1,4) with a neighbouring voltage probe (2,3). 
This can be achieved, for example, by offsetting the top and bottom probes in non-ballistic devices and it is underpinned by a simple conceptual explanation.
If the top (or bottom) lead couples identically to the left and right leads, its potential must lie half way between them $V_{2(3)} = \frac{V_1 - V_4}{2}$. 
Moving the top probe nearer to either the left or right lead changes its potential, and breaking the symmetry between top and bottom probes then gives a finite $V_{23}$.

In homogeneous spin-orbit systems, with a symmetric placement of leads, we assume that the transmissions between neighboring leads depends only on their relative alignment, i.e. clockwise (c.) or anti-clockwise (a.c.), and on the spin orientation. 
Using the identity $T_{ \alpha \beta}^\downarrow = T_{\beta \alpha}^\uparrow$ to relate the two spin channels with SOC, we write
\begin{equation}
	\begin{aligned}
	T_{42}^\uparrow &= T_{34}^\uparrow = T_{13}^\uparrow = T_{21}^\uparrow  \\
      & = \; T_{24}^\downarrow = T_{43}^\downarrow = T_{31}^\downarrow = T_{12}^\downarrow \equiv T_{\mathrm{a.c.}}^\uparrow \equiv T_{\mathrm{c.}}^\downarrow \\
	T_{24}^\uparrow &= T_{43}^\uparrow = T_{31}^\uparrow = T_{12}^\uparrow  \\
      & = \; T_{42}^\downarrow = T_{34}^\downarrow = T_{13}^\downarrow = T_{21}^\downarrow \equiv T_{\mathrm{c.}}^\uparrow \equiv T_{\mathrm{a.c.}}^\downarrow 
	\end{aligned}
	\label{homoSOCconditions}
\end{equation}
and quickly find $R_{xy}=0$ from Eq. \eqref{Rxy_simple}.

% (as shown example figure) so that we expect, e.g., $T_{21} > T_{42}$ and $T_{43} >  T_{42}$, and a finite, (positive?) $R_{xy}$.
%This is a simple conceptual explanation for this -- if the top lead couples identically to the left and right leads, its potential must lie half way between them $V_2 = \frac{V_1 - V_4}{2}$. 
%We note that this situation occurs even if one of the voltage probes has a much more stronger couling to the current probes than the other: it is not sufficient to have $T_{21} > T_{31}$ and $T_{42} > T_{43}$ -- we require $T_{21} \ne T_{42}$ or $T_{31} \ne T_{43}$.
%A similar argument holds for the bottom lead, so that the potential at the top and bottom leads is identical and $R_{xy} \sim V_{xy} = 0$.  

No charge current flows in the voltage probes; however a net spin current is possible because spin mixing is allowed in these leads, so that they can absorb and reinject electrons with opposite spin as long as the net charge current is zero.
The general expression for the spin current in each lead is 
\begin{equation}
\begin{aligned}
	I_{\alpha}^s & = I_{\alpha}^\uparrow - I_{\alpha}^\downarrow \\
	& = \frac{e^2}{h}  \sum_{\beta} \left\{ \left( T_{\beta \alpha}^\uparrow - T_{\beta \alpha}^\downarrow\right)  V_{\alpha} - \left( T_{ \alpha \beta}^\uparrow - T_{ \alpha \beta}^\downarrow \right) V_{\beta} \right\} \,,
	\label{eq:Ispin}
\end{aligned}
\end{equation}
which in its expanded form is quite complicated.
For a homogeneous SOC and symmetric leads, using Eq. \eqref{homoSOCconditions}, it simplifies significantly to
\begin{equation}
	I_{2}^s = - I_{3}^s = \frac{V e^2}{h} (T_{\mathrm{c.}}^\uparrow - T_{\mathrm{c.}}^\downarrow) \,
	\end{equation}
-- the hallmark transverse spin current of the SHE.
%The SOC-induced spin Hall effect portrayed here generates a spin current, but not a transverse resistance.
All-electrical detection of the SHE usually employs a non-local setup where this spin current generates a potential difference between an additional pair of probes via the \emph{inverse} Spin Hall effect (iSHE).
A similar effect can be demonstrated in the cross geometry [see \cref{fig:spinsplitting}(e)] by using a ferromagnetic contact as our injector, so that the left lead only injects spin-up electrons, with the other leads unchanged.
Setting $T_{\alpha1}^\downarrow = T_{1\alpha}^\downarrow = 0$ we find
%\begin{equation}
%R_{xy} \sim T_{21}^\uparrow \left(T_{13}^\uparrow + T_{34}^\uparrow + T_{43}^\uparrow\right) - T_{13}^\uparrow \left(T_{12}^\uparrow + T_{24}^\uparrow + T_{42}^\uparrow\right) \,
%\label{Rxy_soc_fm}
%\end{equation}
\begin{equation}
R_{xy} \sim T_{21}^\uparrow \left(T_{13}^\uparrow + T_{43}\right) - T_{13}^\uparrow \left(T_{12}^\uparrow  + T_{42}\right) \,
\label{Rxy_soc_fm}
\end{equation}
for TRS SOC pairs, which simplifies to 
%\begin{equation}
%\begin{aligned}
%R_{xy} & \sim \left(T_{21}^\uparrow - T_{12}^\uparrow \right)  \left(T_{12}^\uparrow + T_{21}^\uparrow \right) \\ 
%& \sim {T_{21}^\uparrow}^2 - {T_{12}^\uparrow}^2
%\end{aligned}
%\label{Rxy_soc_fm_simple}
%\end{equation}
\begin{equation}
R_{xy} \sim {T_{\mathrm{c}}^\downarrow}^2 - {T_{\mathrm{c}}^\uparrow}^2
\label{Rxy_soc_fm_simple}
\end{equation}
in the homogeneous, symmetric case. 
Here the finite potential difference across the device is directly connected to differences between spin-up and spin-down transmission probabilities. 
Swapping the spin-orientation of the injector changes the sign of the transverse resistance, confirming its spin-related origin.
%Thus, signatures of the spin-splitting effect can be captured in the cross geometry through the generation of spin-current in the leads or a transverse resistance in the presence of a polarised injector.

We now turn back to magnetic triangles, which also induce a spatial splitting of spin channels, but obey a different set of symmetries.
As discussed above, each spin channel has a different potential landscape and effectively conserves TRS.
%Firstly, the two spin channels are no longer be simply related to each other due to differing potential profiles.
%Neglecting spin orientation, the Hamiltonian describing each channel consists only of real-valued hopping and on-site terms, and so preserves TRS.
Thus $T_{\alpha\beta}^\downarrow \ne T_{\beta\alpha}^\uparrow$, but $T_{\alpha\beta}^\uparrow = T_{\beta\alpha}^\uparrow$ and $T_{\alpha\beta}^\downarrow = T_{\beta\alpha}^\downarrow$, so that the total transmission also obeys $T_{\alpha\beta} = T_{\beta\alpha}$, despite the absence of overall TRS. 
With unpolarized leads, the analysis based on Eq. \eqref{Rxy_simple} still holds, and the transverse resistance is zero for symmetric probe configurations.
The isotropic condition assumed for SOC cases is no longer a relevant limit here, as geometric anisotropies are essential to the effects studied. 
However, pristine triangles are symmetric upon reflection in the $y$-axis, and asymmetries due to triangle position and disorder should average out in large enough samples, so it is useful to consider the case of left-right symmetric transmissions. 
This forces both $R_{xy}=0$ and $I_{2,3}^s=0$---a critical difference between SOC and magnetic-scattering cases.
For SOC, each spin channel generates an equal and opposite transverse current leading to a finite spin, but zero charge, current, whereas for zz-TGAs each channel independently preserves TRS, so that neither generates a transverse current and both their sum and difference are zero.
Neither scenario generates a non-negligible transverse resistance with non-magnetic leads.
With a FM injector lead however, the zz-TGA system generates a non-zero transverse resistance 
%\begin{equation}
%R_{xy} \sim T_{21}^\uparrow \left(T_{43}^\uparrow + T_{43}^\downarrow\right) - T_{31}^\uparrow \left(T_{42}^\uparrow + T_{42}^\downarrow \right) \,
%\label{Rxy_tri_fm}
%\end{equation}
\begin{equation}
R_{xy} \sim T_{21}^\uparrow T_{43} - T_{31}^\uparrow T_{42}  \,
\label{Rxy_tri_fm}
\end{equation}
which simplifies to 
\begin{equation}
 R_{xy} \sim T_{21}^\uparrow T_{43}^\downarrow - T_{31}^\uparrow T_{24}^\downarrow
 \label{Rxy_tri_fm_simple}
\end{equation}
under left-right symmetry. 
The result from a full simulation of this quantity in a cross-device is shown in \cref{fig:spinsplitting}(f).
Finally, we note that this iSHE-type behavior for zz-TGAs emerges in the absence of a corresponding SHE-type behavior for the charge current.

\section{Conclusion} \label{sec:concl}

We have examined graphene systems patterned with zz-TGAs and demonstrated how large-scale, disordered samples display a range of spatial anisotropies in their charge and spin transport characteristics.
Electronic band-gaps are found for both aligned and randomly-oriented triangles, whereas the aligned case also shows a strong half-metallic behaviour, with up- and down- spin electrons dominating transport at opposite sides of half-filling.
These properties are robust in the presence of strong edge disorder, and in the absence of a superlattice, where they have previously been predicted\cite{Gregersen2016}.

We have further demonstrated that such systems display significant anisotropic transport behaviour with respect to the direction of current flow.
A strong preference for armchair direction transport is found, leading in certain cases to quasi-ballistic behaviour in one direction and localization in the other, despite a homogeneous distribution of perforations.
The qualitative anisotropic trends survive the introduction of edge disorder, suggesting application in realistic devices where a control of the direction of flow of charge current, or spin-polarised current is required.

Triangular perforations can also scatter electrons anistropically in directions perpendicular to the charge current.
The generation of a transverse resistance by a spin-polarised current is the essence of the iSHE, and we note that zz-TGAs can give rise to $R_{xy}$ whose functional form (Eq. \eqref{Rxy_tri_fm_simple}) is similar to that for the SOC case (Eq. \eqref{Rxy_soc_fm_simple}).
Furthermore, we note that, for zz-TGAs, the iSHE-type behaviour emerges in the absence of a corresponding SHE-type behaviour for the charge current and zero bulk transverse resistivities.
Therefore zz-TGA devices offer the possibility not just to filter different spin channels into different leads, as discussed in Ref. [\onlinecite{Gregersen2017}], but of generating a transverse resistance whose sign depends on the spin-polarisation of the input current.

A key feature of zigzag-edge magnetism in graphene is the spin-dependent asymmetry with respect to Fermi energy, so that sweeping through half-filling inverts the role of the two spin channels.
This leads to an antisymmetry in the $R_{xy}$ measured with an FM injector, as the spin-polarised current is deflected towards opposite edges of the system for different charge carriers.
The symmetry with respect to $E_F$ in standard iSHE setups depends on the type of SOC considered: intrinsic Kane-Mele coupling should present a symmetric signal\cite{PhysRevLett.117.176602} whereas Rashba coupling should give an asymmetric signal with a linear transition\cite{chen2012spin, PhysRevLett.117.176602} through $E_F=0$.
This suggests that contributions from zigzag-edge magnetism could result in Rashba-like behaviour being observed in inverse Spin Hall measurements.

Finally, the behaviors discussed here emerge from the presence of two key features in our systems: i) local magnetic defects with a sublattice-dependent antiferromagnetic alignment and ii) geometric asymmetries.
These features arise naturally for zz-TGAs, but we note that it may be be possible to generate similar effects, deliberately or inadvertently, in other systems.
A number of individual defects, including vacancies, hydrogen adatoms and substitutional species, are predicted to induce local magnetic moments in graphene.
The alignment of these moments, through e.g. indirect exchange interactions\cite{Saremi2007, Power2013}, is expected to be highly sublattice-dependent. 
Local clustering, or an asymmetric occupation of sublattices, could give rise to the conditions necessary for anisotropic transport or transverse resistance signals.
This is perhaps most relevant for hydrogen atoms, which have been proposed as both a source of local magnetism\cite{PhysRevB.75.125408, gonzalez2016atomic} and of spin-orbit coupling\cite{PhysRevLett.110.246602}.
Experimental signatures of SHE and iSHE mechanisms have been detected\cite{balakrishnan2013colossal}, but have also proven difficult to reproduce\cite{PhysRevB.91.165412}, suggesting that a more complex mechanism than a simple enhancement of SOC is at play\cite{kochan2014spin, soriano2015spin}. 
In this direction, a recent study\cite{2018arXiv180107713O} has also highlighted the importance of the interplay between magnetism, disorder and SOC in determining the exact Hall response in graphene systems.

\section{Acknowledgments} \label{sec:ackno}
S.S.G. acknowledges research travel support from Torben og Alice Frimodts Fond and Otto M{\o}nsted Fonden.
S.R.P. acknowledges funding from the European Union’s Horizon 2020 research and innovation programme under the Marie Skłodowska-Curie grant agreement No 665919. 
J.H.G and S.R. acknowledge funding from the Spanish Ministry of Economy
and Competitiveness and the European Regional Development
Fund (project no. FIS2015-67767-P MINECO/FEDER, FIS2015-64886-C5-3-P)
and the European Union Seventh Framework Programme under grant agreement
no. 785219 (Graphene Flagship).  
The Center for Nanostructured Graphene (CNG) is sponsored by the Danish Research Foundation, Project DNRF103.
ICN2 is funded by the CERCA Programme/Generalitat de Catalunya and supported by the Severo Ochoa programme (MINECO, Grant. No. SEV-2013-0295).

\end{document}